The COVID-19 pandemic's effects on poor rural dwellers in sub-Saharan Africa: A case study of access to basic clean water, sanitary systems and hand-washing facilities

[1]John Stephen Kayode*, [2]Asha Embrandiri, and [3]Adijat Olubukola Olateju

[1]Institute of Hydrocarbon Recovery, Department of Research and Innovation. Universiti Teknologi PETRONAS, Persiaran UTP, 32610 Seri Iskandar, Perak Darul Ridzuan, Malaysia.

[2]Department of Environmental Health, College of Medicine and Health Sciences Wollo University Dessie, Ethiopia

[3]Department of Economics, Faculty of Social Sciences, Lagos State University, Ojo Campus, Lagos State, Nigeria.

*Corresponding Author: jskayode@gmail.com, john.kayode@utp.edu.my

Abstract

The fear of the invisible but prevalent Coronavirus (COVID-19), disease cannot be overemphasized since there is the potential possibility of it wiping out the entire world population within a few months if adequate and quick steps are not taken to curb this menace, and the sub-Saharan African (SSAn) region is no exception. It is evident that water, as an essential daily commodity, has long been in a state of emergency in SSAn nations, which is largely attributed to decades of neglect by the successive governments, because it has not been possible to separate the existing bond between water, health, livelihood and the economy. The laudable Millennium Development Goals (MDGs) proposed by the United Nations had yet to achieve the stated objective of improving the standards of living and health conditions of the rural communities in the SSAn region before the COVID-19 pandemic outbreak. This failure has been masked by a sort of delusion in which the people of this region are subjected to the hardship of searching for clean and healthy water in their own ponds, rivers, streams and shallow hand-dug local wells on a continuous basis. Less than 17% of the rural population in all the SSAn communities can access basic hand-washing facilities and sanitation systems. The total water productivity, as measured by the Gross Domestic Product (GDP) per cubic meter of total freshwater withdrawn, for the people was <5 GDP.

1. Introduction

Prior to the outbreak of the COVID-19 pandemic, the World Economic Forum categorized water scarcity as the third most alarming global risk. Sub-Saharan Africa (SSA) is composed of 46 countries in Africa. Its erratic and low amount of a good and clean water supply predates the colonial era, and many communities suffer from a lack of access to clean and improved sources of sustainable drinking water. Water and sanitation facilities are absent in many of the rural communities despite some laudable water and sanitation policies by successive governments after the colonial era. A majority of the rural communities in this part of the world are conditioned to the seasonal supply of water resources despite the high water availability during the rainy season (Figure 1). It is no stretch to say that the majority of the governments at all levels failed to make provisions to establish sustainable rural water supply as a priority project to serve the communities. During the prolonged dry seasons that usually characterize the region, which sometimes last for more than half of the year in many places (Figure 1), streams and rivers are completely dried up with many of the hand-dug wells having short supplies due to their shallow nature. The number of wells serving a community is more than what the groundwater situation could handle in many of the places, although the United Nations has invested heavily in the provision of sustainable, clean and safe drinking water in the whole African continent since the promulgation of the millennium development goals (Sachs, 2005, United Nations, 2007, Hunter, 2009). It is unfortunate that most of the rural dwellers have not yet felt the impact, as many of the installed water systems are not functional, and those that were initially operational were left unmaintained after breaking down. There had been no serious policy to improve the access of rural dwellers sustainable and clean water in all the countries of the SSA. As a matter of urgency, the provision of good and safe drinking water in recent years has experienced consistent decreases due to population explosions in the region (UNESCO, 2006, Smeets, 2008, United Nations, 2012).

Previous works have shown relationships between improved clean and good sources of drinking water and drastic reductions in health risks (Bartram and Cairncross, 2010, Brown, et al., 2008, GSunny, et al., 2020, Yang, et al., 2020). The laudable Millennium Development Goals (MDGs) proposed by the United Nations had yet to achieve the stated objective of improving the standard of living and health conditions of the rural communities in the SSAn region before the COVID-19 pandemic outbreak. Hence, the people of the region have been confronted by the menace of the new invisible, yet powerful, Coronavirus disease. If the fight against this deadly pandemic global disease is to be successful in the region, quick attention must be given to the provision of good, clean and safe water, especially for the rural communities that have suffered so much neglect from the governments. It is noteworthy that the various restrictions imposed in the aftermath of the COVID-19 outbreak by the Centers for Disease Control (CDC) would be insignificant in the region as long as people have to travel long distances during the February-April dry period in the area in search of clean, safe and drinkable water. Hence, exposure to the deadly infectious virus cannot be ruled out. The health of the rural dwellers in SSA is more threatened as the people become more vulnerable to getting infected by the deadly disease in their search for healthy, clean, and safe water. The situation became worse due to the non-availability of protective devices such as face masks to protect people, even though people, especially women and children, cannot be prevented from gathering in places where water is available, and safe health cannot be guaranteed. Conversely, the people are exposed to the danger and challenges posed by the pandemic, which could be a channel through which people innocently get infected. Similarly, people are warned to maintain good hygiene through regularly washing their hands with soap, which is the only available means to these rural communities. Unfortunately, the issue of the immediate availability of sustainable water will impair this. The more reasonable information of this nature contained in this paper is imperative since it is the medium for disseminating the required knowledge.

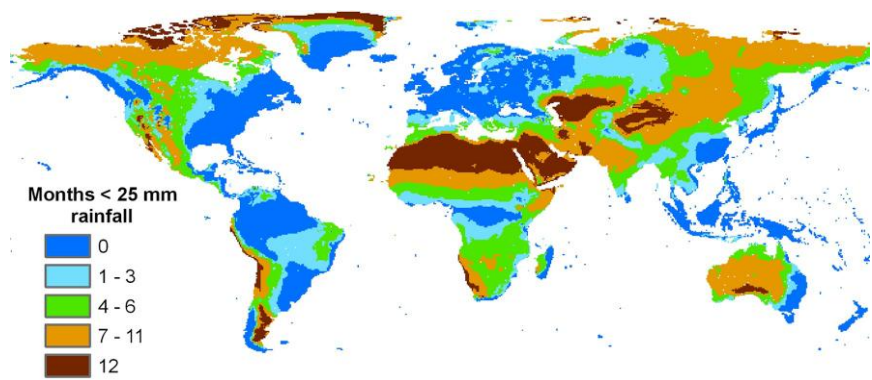

Figure 1: Global Rainfall Distribution, adapted from New, et al., 2000.

Therefore, the provision of improved clean, potable, and safe water and sanitary facilities in the SSA calls for the declaration of a state of emergency regarding the region's water supplies if the global fight against the COVID-19 pandemic can be won as soon as possible. Good health cannot be obtained in the absence of affordable, easily accessible, safe, and dependable potable water for the rural communities of SSA. The respective governments' attention needs to be refocused on the nonfunctional water facilities in many places, such as boreholes and hand-dug wells that were abandoned and left unmanaged in numerous places, either as a result of the lack of constant power supplies, or the lack of maintenance. In the present uncertain situation surrounding the COVID-19 pandemic, it is critical for the people to get access to clean and safe sustainable water for domestic and sanitation uses to guarantee public hygiene and better food and, hence, improve their health conditions (Hunter, et al., 2009).

Long before the COVID-19 crises started, the provision of clean and safe water for the use of rural communities in SSAn nations saw the burden shifted to the poor rural community dwellers by the successive political governments. This burden has been masked by a sort of delusion in which the people are subjected to the hardship of searching for clean and healthy water in their own in ponds, rivers, streams and shallow hand-dug local wells on a continuous basis. It is unfortunate that the depths to the permeable underground water table in the ground areas of the region are too shallow to provide water throughout the year (Jha, et al., 2020, Kayode et al., 2016), given the complexity of the topography and the effect of the extreme seasonal weather conditions in the region caused by the prolonged dry periods that subject these shallow wells and the streams to a few months of complete dryness. These situations in SSA have a great effect on the provision of sustainable, clean, and safe water for both agricultural and domestic use.

**2. Methodology**

2.1 COVID-19 pandemic and the rural communities of sub-Saharan Africa (SSA)

The COVID-19 pandemic has caused global public health crises across entire regions of the world (Figure 2) with no exceptions. It was envisaged that the situation in SSA, and Africa as whole, would be devastating according to the WHO Regional Director for Africa (RDA) Dr. Matshidiso Moeti in a 7th April 2020 report (WHO, 2020a). In the RDA's report, the disease was first discovered in the north African country of Egypt on 14th February 2020. As of 30th April, all African countries had recorded a COVID-19 case because of the influx of the people from the earlier hotspot zones in Asian and European countries, as well as the United States, according to Dr. Moeti. Furthermore, the possibility of a pandemic outbreak involving family clusters could not be completely ruled out after the discovery of the first case (Guan, et al., 2020, Phan, et al., 2020). Most reports have shown the increased risks of people with chronic health conditions of getting infected by COVID-19, particularly adults (Guan, et al., 2020, La Rosa, et al., 2020, Phan, et al., 2020, Shereen, et al., 2020, Wang, et al., 2020, WHO, 2020b). The risk of getting infected by individuals with some prior health conditions is very high.

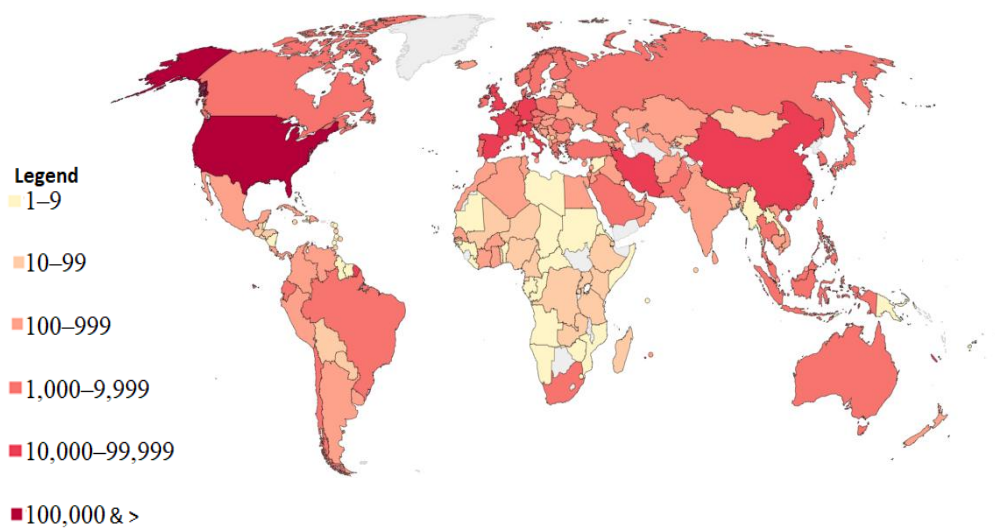

Figure 2: Confirmed COVID-19 cases by countries as of 30/04/2020. Source: Corona-Statistics-And-Tracker-Dashboard-Angular-9 (Rafique, 2020).

If the projected impacts of the COVID-19 pandemic on the people of SSA are to be believed, then the major gains in the sustainability and availability of potable water for these people would suffer a drastic set back, including the areas supposedly known as urban cities. Situational reports in earlier studies suggested declaring a state of emergency regarding the provision and supply of potable water to meet the daily needs of the increasing population in the region. The general state of the potable water supply in many of the cities is worrisome. The exponential increase in the cases of COVID-19 recorded so far makes it of utmost importance to research the major impact that the clinical disease would have on the SSAn poor rural communities that could not access potable and sustainable water, coupled with the growing population and the increase in the demand for sustainable and potable clean water due to the increasing economic constraints facing the people.

**3. Discussion of results**

As of the end of 2018, the World Bank Group reported the mortality rates in SSA that are presented in Figure 3. It is alarming to see a whopping percentage of a little less than a 50% of every 100,000 people of the region losing their lives to unsafe water, sanitary systems, and unhygienic conditions. The rapid spread of the COVID-19 pandemic is a serious concern when considering the data from the World Bank Group's reports on SSA. The results of the mortality rates presented in Figure 3 clearly show that the deaths due to unsafe water, the complete absence or improper sanitary systems, and unhygienic conditions are more than double the deaths from every other source put together. The unfortunate situation is the lack of political will on the part of the governments of the countries that make up the SSAn region vis-a-vis proactively dealing with the situation of sustainable rural water supplies and the water systems' management over several decades of neglect.

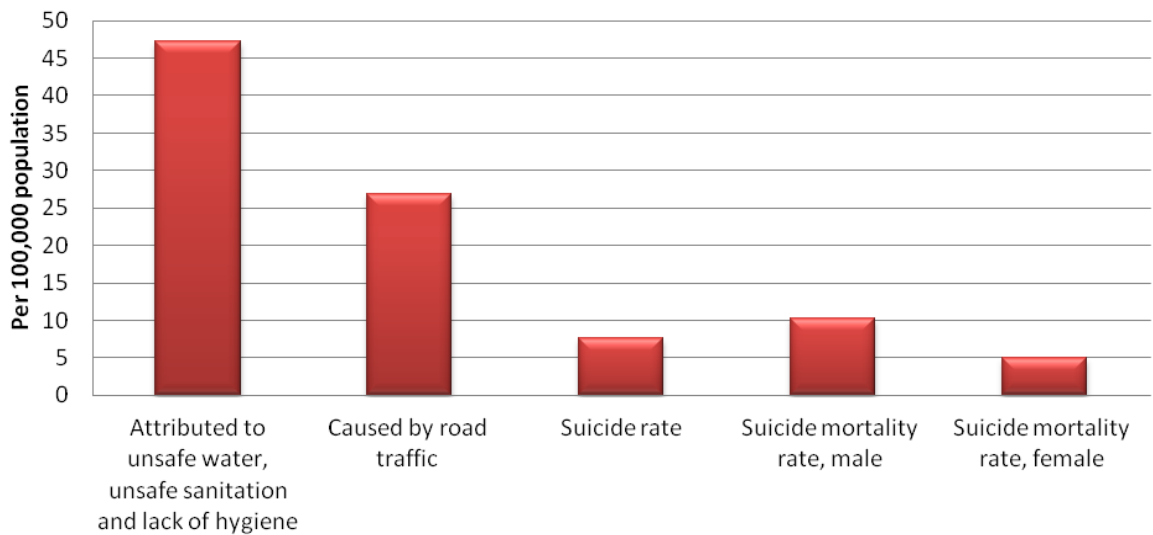

Figure 3: Mortality rates in SSA as of the end of 2018. Data Source: The World Bank Group, 2019.

There have been concerted efforts to establish national water policies across the region, either by the United Nations sanitary projects or by the individual governments in each of the countries that constitute the SSA. However, the absence of sufficient potable water among a larger percentage of the communities is worrisome. Figure 4 shows the percentage of SSAn people with access to basic hand-washing facilities from 2009-2017 as compiled by the World Bank Group reports. The results of the data analysis call for urgent attention if the fight against the COVID-19 pandemic is to be globally shared and fully achieved. Less than 17% of the rural population in all the communities can access basic hand-washing facilities. The percentage was less than 10% in 2013 and jumped by over 8% to the present level of 17% in 2014 and has since then remained at this level. This could be attributed to the concentration of the UNICEF programs in the aftermath of the deadly Ebola virus outbreak. As of the time the data were collected, only <37% of the urban population had access to basic hand-washing facilities. Again, this level of access was made possible because of the Ebola cases. Prior to the Ebola outbreak in the region, barely 19% of the urban population was recorded to have access to basic hand-washing facilities from 2009 to 2011. The percentage was substantially improved from 2012-2013 to about <22%. Overall, only approximately 25% of the entire SSAn population had access to basic hand-washing facilities from 2014 until 2017, leaving a whopping 75% of the population of SSAn without access to basic hand-washing facilities. Due to the necessity of basic hand-washing facilities due to the COVID-19 pandemic, SSAn Nations would need to seriously brace for the challenges brought about by the emergency created by the deadly pandemic. This report is a total reflection of what is obtainable since many water-constrained factors have led to the low level of accessibility to potable water. These factors together with the population explosion of the region over the past decades have increased the demands for accessible and sustainable good drinking water and proper sanitary systems.

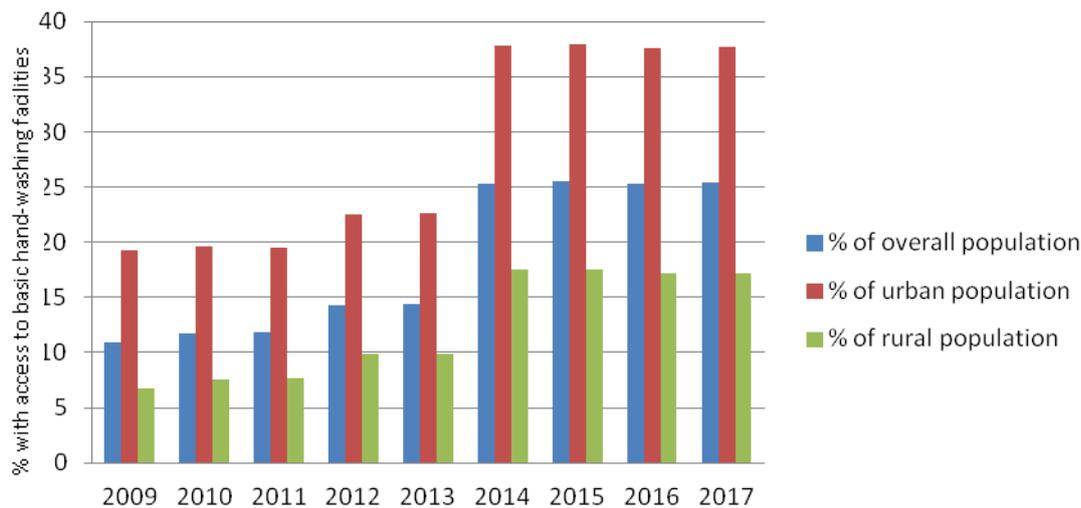

Figure 4: Percentage of people with access to basic hand-washing facilities (i.e., soap and water): (a) overall population, (b) urban population, and (c) rural population. Data from 2009-2017. Source: The World Bank Group, 2019.

There has not been any confirmation so far of people contacting the coronavirus through water bodies, either in underground or surface waters, or spreading it through any contaminated drinking water system (Casanova, et al., 2020, Shereen, et al., 2020, Yang, et al., 2020), despite the virus being detected in all types of water systems, with a reported connection between drinking and recreational water outbreaks. The provision of sustainable, clean, safe, and potable water for domestic and sanitary systems should be pursued vigorously for the SSAn people. In the absence of this pursuit, a second wave of the disease could outbreak and spate the area with unforeseen devastating effects that increase global concerns. This paper therefore provides proven knowledge and valuable support for sustainable, hygienic, clean, safe water and good sanitary systems for the rural poor dwellers in the SSA to safeguard the people from unrestricted exposure to possible infection from the COVID-19 disease through water and sanitation systems, with the level of cases recorded as at 30th April shown in Table 1. People cannot maintain good sanitation and hygienic conditions in situations where the basic water supply required to do so is completely absent. The only avenue for efficiently enhancing public health in rural communities relies solely on the provision of adequate, hygienic, clean and safe water and good quality sanitary systems. The provision of clean and quality domestic water supplies is the key to maintaining good health and, hence, preserving lives.

Table 1. Confirmed Cases by Countries in the SSA as at 30th April 2020. Data source: Corona-Statistics-And-Tracker-Dashboard-Angular-9 (Rafique, 2020).

| Country | No of Cases |
|---|---|
| South Africa | 5,951 |
| Nigeria | 2,170 |
| Cameroon | 1,832 |
| Ghana | 2,074 |
| Guinea | 1,537 |

Table 1. Continue

| Country | No of Cases |
|---|---|
| Cote d'Ivoire | 1,333 |
| Senegal | 1,024 |
| Niger | 728 |
| Burkina Faso | 649 |
| Congo (Kinshasa) | 604 |
| Somalia | 601 |
| Sudan | 533 |
| Mali | 508 |
| Tanzania | 480 |
| Kenya | 411 |
| Mauritius | 332 |
| Equatorial Guinea | 315 |
| Gabon | 276 |
| Guinea-Bissau | 257 |
| Rwanda | 249 |
| Congo (Brazzaville) | 229 |
| Liberia | 152 |
| Sierra Leone | 136 |
| Ethiopia | 133 |
| Madagascar | 132 |
| Togo | 123 |
| Cabo Verde | 122 |
| Zambia | 109 |
| Benin | 90 |
| Uganda | 85 |
| Mozambique | 79 |
| Chad | 73 |
| Central African Republic | 72 |
| South Sudan | 45 |
| Zimbabwe | 40 |
| Eritrea | 39 |
| Malawi | 37 |
| Angola | 30 |
| Botswana | 23 |
| Namibia | 16 |
| Sao Tome and Principe | 16 |
| Gambia | 12 |
| Burundi | 11 |
| Seychelles | 11 |
| Mauritania | 8 |
| Comoros | 1 |

Figure 5 shows the results for the total water productivity as measured by the Gross Domestic Product (GDP) per cubic meter of total freshwater withdrawn using data from 1999 to 2016 on the SSAn region from the World Bank Group 2019 report. Ever since the 1999 production of approximately 63 GDP of total water per cubic meter in the SSAn region, it is worrisome that it has seen a drastic freefall from this level of attainment to just approximately 12 GDP in the following year, with a consistent further plunge to a lower value of approximately 5 GDP in 2002. There was a little improvement in the total water productivity for the people in SSA in the following year (i.e., 2003) to about <11 GDP. The next year, 2004, witnessed another sharp drop to <3 GDP, with a record improvement of <15 GDP in the year 2005. However, 2006 experienced a catastrophic level of water productivity in this region, as shown by the drastic drop again to the lowest total water productivity of <2 GDP. The years 2008, 2010, and 2013 experienced consistent improvements, even though the recorded total water productivity was <30 GDP, but these levels have since fallen again to <5 GDP per cubic meter of total freshwater withdrawn in 2016 as obtained from the World Bank Group 2019 report.

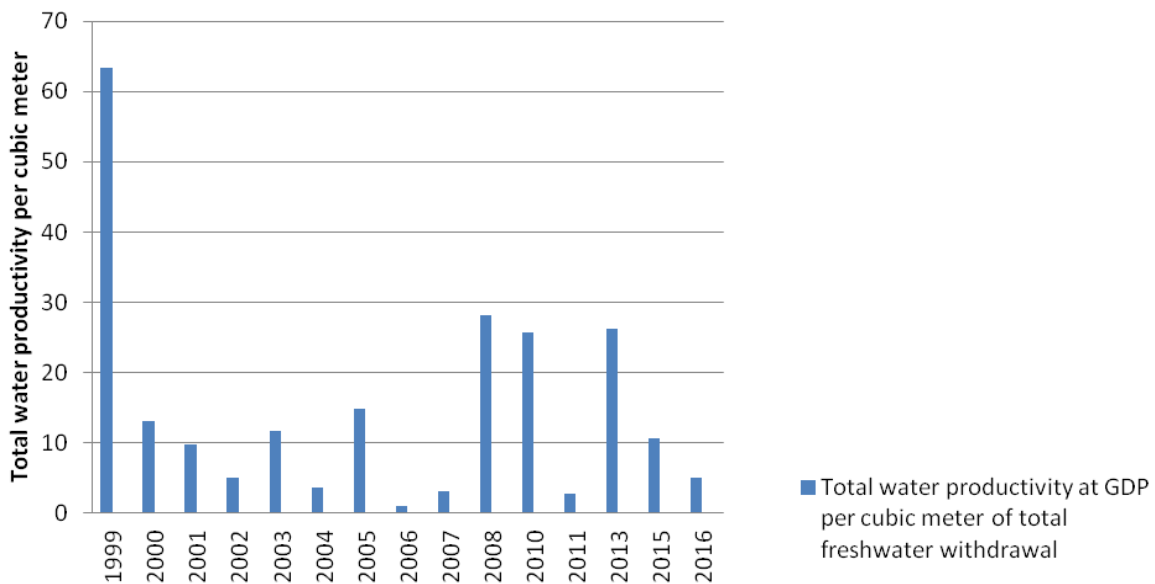

Figure 5: SSAn total water productivity, as measured by GDP per cubic meter of total freshwater withdrawn, using data from 1999 to 2016. Source: The World Bank Group, 2019.

The connections among the major parameters expressed in this study are the following: i) the percentage of people with access to basic hand-washing facilities; ii) the total water productivity, as measured by GDP per cubic meter of total freshwater withdrawn; and iii) the COVID-19 pandemic that has had significant effects on SSAn's rural communities in relation to their social economic status. The decades of experience in real daily life with nonexistent accessibility to sustainable, clean, and potable water for domestic and sanitary systems motivate this paper to publicize the major challenges confronting the people of the region that have defied lasting solutions to attract probable lasting solutions in the face of the COVID-19 pandemic outbreak. The present total water productivity, as measured by GDP per cubic meter of total freshwater withdrawn, across the region needs to be boosted by multiple folds due to the pressure of the demands of the COVID-19 pandemic that has further strained the present insufficient water supply systems. Although concerted efforts have been made to ward off the deadly virus within a short time frame, attention needs to be focused on the SSAn water supply systems. The fight against the COVID-19 pandemic could be a straightforward effort in areas where there are no restrictions to sustainable, hygienic, clean, safe water and sanitary systems. However, it could be much more complicated in the SSAn region since the services of this nature are utterly deficient over long decades in the rural areas with many of the urban cities being inclusive. In a situation where the entire total water productivity, as measured by GDP per cubic meter of total freshwater withdrawn, was <5 GDP in 2106 for the region, there are serious concerns for these groups of people due to the level of hygiene required to face the novel coronavirus diseases. The most effectual and clear-cut measures to thwart the transmission of this deadly disease is to maintain high-quality, hygienic conditions through unrestricted access to sustainable, hygienic, clean, safe water and good sanitary systems.

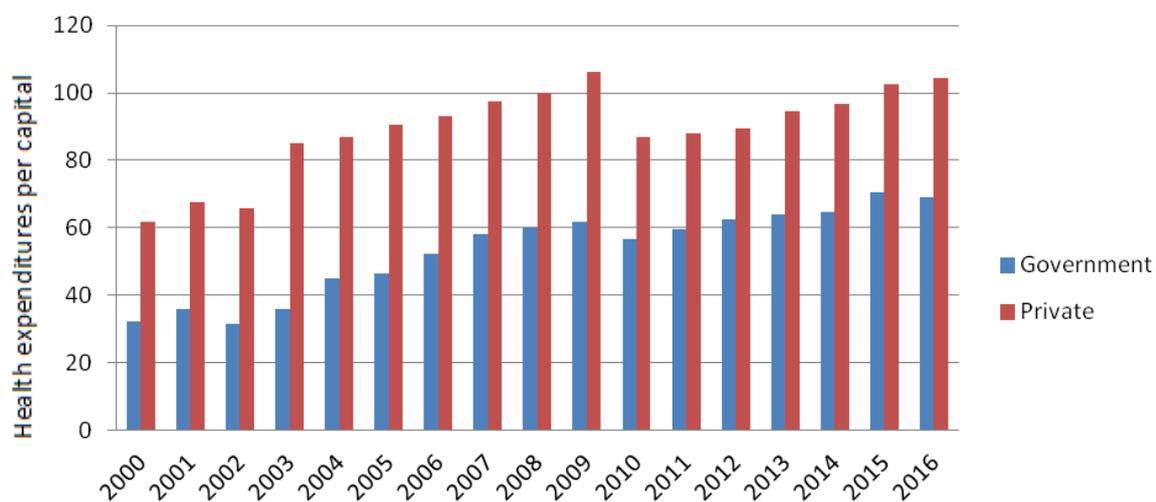

Figure 6: SSAn health expenses per capita (comparing the government commitment with what individuals/private people spent on health care over 17 years). Source: The World Bank Group, 2019.

3. Sub-Saharan African health care systems and the COVID-19 pandemic.

We compare the SSAn governments' commitment to the health care systems with what individuals or private people spent on health care across 17 years using data obtained from the World Bank Group 2019 report and presented it in Figure 6. After a close examination of the trends, it is very clear that the governments of the nations in the SSAn region paid less attention to the health care systems of the people, particularly the rural dwellers. Individuals have been left to provide health care for themselves and possibly their immediate family members. The latest rural community population was 645,077,217 million as of the end of 2018 according to the population data obtained from the World Bank Group 2019 report (Figure 7). These people, for a large extent of the SSAn rural population, are at the highest risk of being easily exposed to the deadly novel coronavirus disease. This risk stands despite the fact that the region continuously suffers from the effects of other deadly global infectious diseases that were long ago found to be closely connected to the non-availability of a sustainable water supply and hygienic sanitary systems. The communities in the region would definitely experience severe detrimental effects as the COVID-19 pandemic continues to ravage everywhere with no exception for any region.

The urban population of 433,229,303 million people in SSAn is not spared from the risks of the COVID-19 pandemic, and most of the cases so far have been reported in urban cities. The percentage of people with access to basic hand-washing facilities, coupled with the available total water productivity, as measured by GDP per cubic meter of total freshwater withdrawn, cannot adequately care for these populations of people. Many of the diseases associated with non-accessibility to hygienic water are very much prevalent in urban cities compared to rural areas.

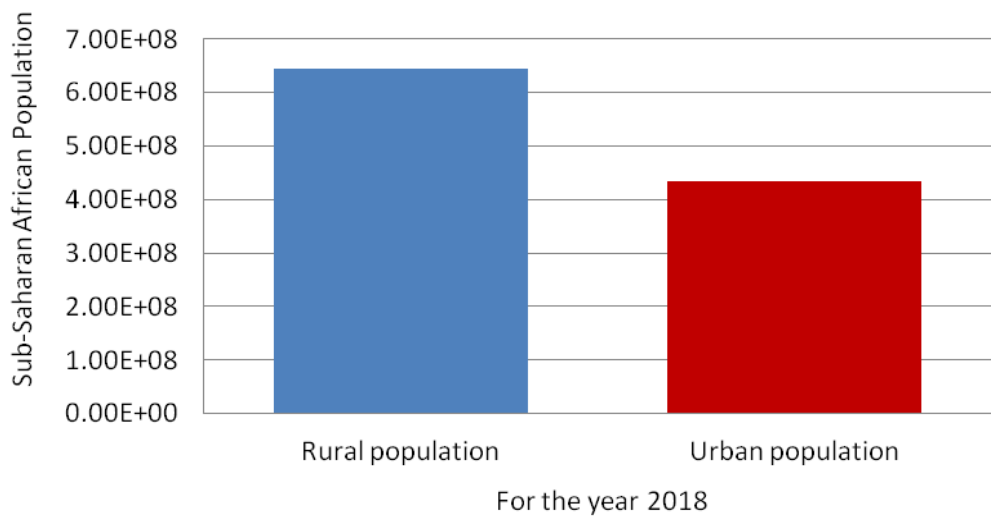

Figure 7: 2018 SSAn total population. Source: The World Bank Group, 2019.

Presently, there is no existing known vaccine for COVID-19 worldwide. However, scientists are vigorously working on remedies that will effectively reduce the number of novel coronavirus infections. While the best pathogenesis and treatment alternatives are still been pursued globally (Guan, et al., 2020, La Rosa, et al., 2020, Phan, et al., 2020, Shereen, et al., 2020, Wang, et al., 2020, WHO, 2020b), researchers are putting their heads together to find ways of reducing the spread of COVID-19. The concerns for the rural dwellers in the SSAn region is the ability to enact with the solutions defined for the novel coronavirus. The economic conditions of this group of people would definitely subject them to be more vulnerable to the infectious disease, especially due to the non-accessibility to basic daily needs such as good food; sustainable, hygienic, clean, and safe water; and good sanitary systems. The people are still suffering from the other deadly waterborne infectious diseases that are continuously ravaging the region on a continuous basis. The outbreak of the COVID-19 pandemic would definitely compound the issues of immunities and malnutrition with the people of these communities.

Figure 8 showed the 2018 percentage of the total population and growth data for the SSAn region, which was obtained from the World Bank Group 2019 report. Rural areas account for 56% of the entire population, with a growth rate of just 2% due to the rural-urban migration of the people in search for better living conditions. Although the urban population is just 38% in the region, the growth rate of 4% is a concern since the mingy available amenities are under stress on a continuous basis as the people scramble for the little available sustainable water supply and hygienic sanitary systems.

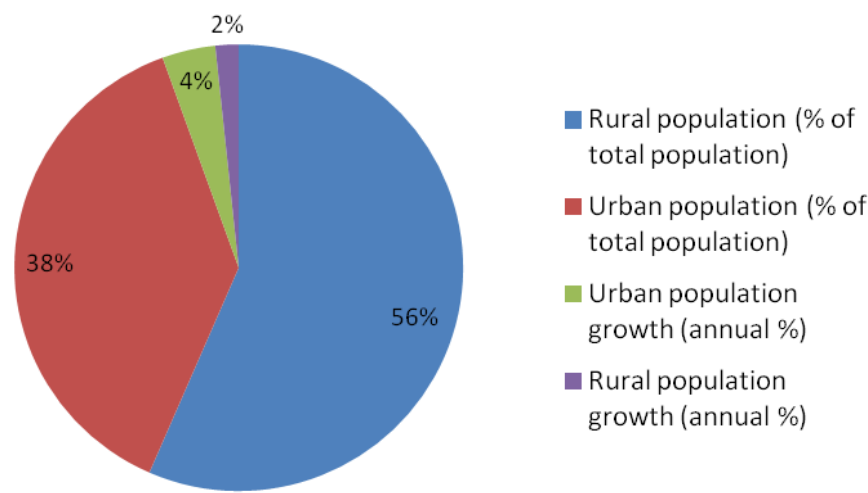

Figure 8: 2018 percentages of the total population and growth in SSAn. Source: The World Bank Group, 2019.

The large population of people living below $1.90 per day (i.e., Figure 9) would not be able to afford the minimum costs of COVID-19 sanitizer that has an average cost of $1.89 per unit. It is certain that the population of people living below $3.20 and $1.90 cannot afford the minimum costs of the basic supplies to fight against COVID-19. It should pointed out at this juncture that many of the statistics made available by the individual nations in this region may perhaps significantly vary from the World Bank Group reports since the governments in most of the countries have significant corruption that politically undermines the true realities of the situations. The situations are extremely horrific with no exception for the urban dwellers too. Figure 10 shows the proportion of the SSAn population in percentages living below the 2011 World Bank poverty line of $1.90 a day and the $3.20 poverty line that accounts for health care expenditures.

The fear of this invisible but prevalent Coronavirus disease cannot be overemphasized since there is the potential possibility of it wiping out the entire world population within a few months if adequate and quick steps are not taken, and the SSAn region is no exception. These enlightening data should focus the attention and concerns of policy formulators on the need to urgently improve and increase the number of new techniques to improve the water supply situation in the region. It is evident that water, as an essential daily commodity, has long been in a state of emergency in SSA, which is largely attributed to decades of neglect by governments, because it has not been possible to separate the existing bond between water, health, livelihood and the economy. All these go pari passu.

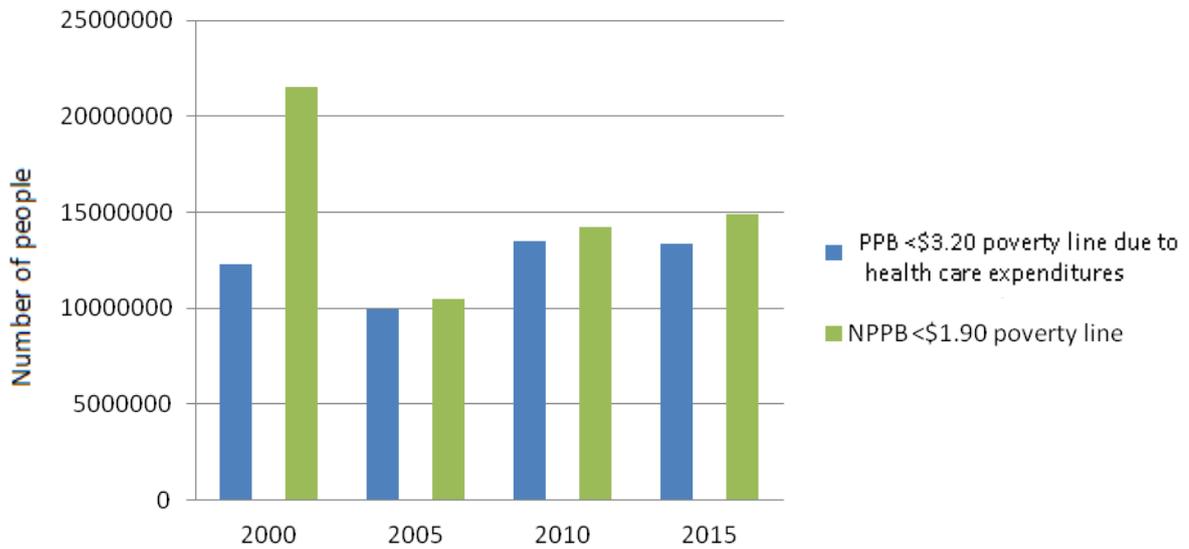

Figure 9: Number of people in SSAn living below the 2011 World Bank poverty line due to health care costs per capita. (a) people living below $3.20, and (b) people living below $1.90. Source: The World Bank Group, 2019.

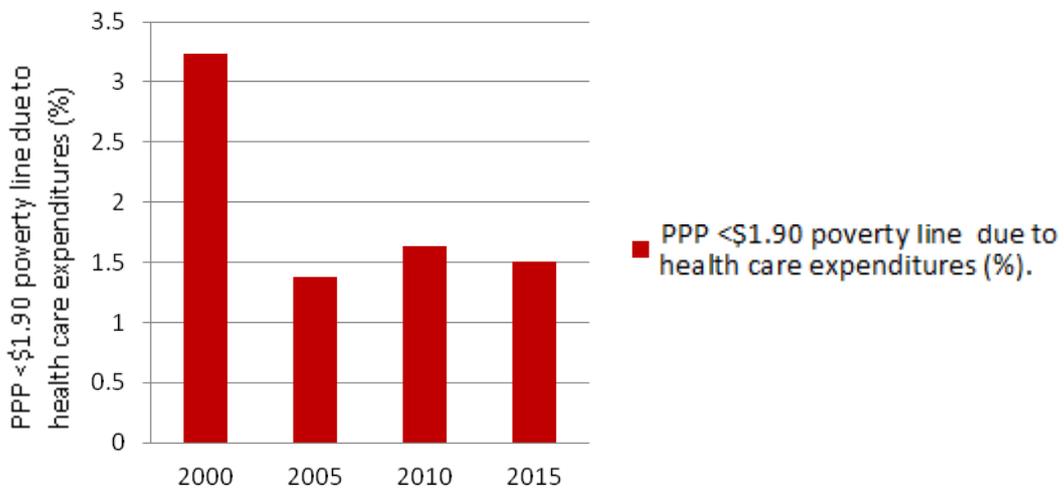

Figure 10: Proportion of the SSAn population in percentage living below the 2011 World Bank poverty line of $1.90 a day due to health care expenditures. Source: The World Bank Group, 2019.

4. Conclusion

The actual reality of the day-to-day lives in SSAn nations is clearly shown by this study. If the necessary efforts had been made to improve the access to sustainable, good and clean water in these developing countries, the whopping >47 million deaths due to inappropriate water, sanitation and hygienic conditions that were presented in Figure 3 could have been prevented. The actual situations in virtually all the countries are heavily characterized by corruption in all spheres of the economy that politically undermines the true reality of the situation. The people would definitely be subjected to further vulnerability to this infectious disease due to their mingy economic conditions due to the non-accessibility to basic daily needs such as good food; sustainable, hygienic, clean, and safe water; and good sanitary systems. The situations being reported are not limited to rural dwellers, and they exist in all the countries comprising the SSA. Challenges of access to clean and safe water together with good sanitary systems predate the Colonial regimes when there were far fewer people than what exist today. Decadent structural facilities are ubiquitous in this part of the globe. The entire blame should be upon the political class that fails to provide basic social amenities that are crucial to life such as clean and good water supplies. Despite the many government policies that address the supply of hygienic, clean, and safe water in most places, the results presented in Figure 5 confirm that the water supply systems are quickly deteriorating at astronomical rates. Virtually all the countries in this region enjoyed numerous World Bank water projects, but the true situation is that the rapid population explosion has jeopardized all these laudable programs. The results shown in Figure 10 regarding the population in percentages that lived below $1.90 a day in 2015 imply that over 16 million of the population lived below $1.90 a day. The figure could be double this when new reports are published.

The outbreak of the COVID-19 pandemic could be judged to have originated from poor risk management on the part of health scientists that should have proffered solutions at the early stage. More worrisome is the unprepared research and academic institutions not curbing the astronomical spread of the coronavirus disease. The policies that are responsible for implementing sustainable, hygienic water and sanitary systems in the region were at significant risk for decades before the COVID-19 pandemic issues. For instance, since 1999 when the GDP production of total water per cubic meter recorded in the SSAn region stood at 63 GDP (i.e., Figure 5), the measure has experienced a drastic free fall to <5 GDP in 2016, which is an indication the water supplies have reached a state of emergency. This indication may not be unconnected to the consistent rise in people's health costs per capita, as presented in Figure 6. It could be concluded that this growth was a result of water supply crises coupled with growing and competing demands at the expense of non-existing sustainable, clean and hygienic water and sanitation resources in several places. With the water situation in SSA having a strong association with the outbreak of many known diseases, such as Ebola, Malaria, and other tropical infectious diseases, this paper suggests the most important and intensive efforts for improving water and sanitation in order to achieve the total and complete eradication of these deadly diseases as urgently as possible. To be equipped to meet the United Nations Sustainable Development Goals (SDGs) regarding targets for the total eradication of these infectious diseases through sustainable water and sanitation programs, a stern obligation of the political governments of each country in the SSAn region is to guarantee the people, especially the rural dwellers, unrestricted access to basic daily needs of sustainable, hygienic, clean, and safe water and good sanitary systems.

**Acknowledgments**

We thank the World Health Organization for the release of the 2020 reports where a larger percentage of the data were obtained. Thanks are also due to ESRI for the access to the COVID-19 Live Tracker.

**Competing Interests**

The authors declare that they have no known competing financial interests or personal relationships that have, or could be perceived to have, influenced the work reported in this article.

**Author Contribution Statement**